# Systematic Analysis of Programming Languages and Their Execution Environments for Spectre Attacks

Amir Naseredini[1,3], Stefan Gast[2,3], Martin Schwarzl[3], Pedro Miguel Sousa Bernardo[4], Amel Smajic[3], Claudio Canella[3], Martin Berger[1,5], Daniel Gruss[2,3]

[1] *University of Sussex, United Kingdom*

[2] *Lamarr Security Research, Austria*

[3] *Graz University of Technology, Austria*

[4] *Instituto Superior Técnico, Universidade de Lisboa, Portugal*

[5] *Turing Core, Huawei 2012 Labs, London, UK*

Abstract:
In this paper, we analyze the security of programming languages and their execution environments (compilers and interpreters) with respect to Spectre attacks. The analysis shows that only 16 out of 42 execution environments have mitigations against at least one Spectre variant, *i.e.*, 26 have no mitigations against any Spectre variant. Using our novel tool Speconnector, we develop Spectre proof-of-concept attacks in 8 programming languages and on code generated by 11 execution environments that were previously not known to be affected. Our results highlight some programming languages that are used to implement security-critical code, but remain entirely unprotected, even three years after the discovery of Spectre.

## 1 INTRODUCTION

In conditional branch instructions, deciding whether to branch or not requires evaluating the branch condition, which can depend on a slow operation, such as main memory access. In most programs, branches are highly predictable from information available before the branch condition has been evaluated, primarily past outcome of related conditional branches. This predictability is exploited using branch predictors, adaptively trained as the CPU executes, on past branching behavior, see [Mittal, 2018] for an overview. Branch prediction allows CPUs to execute the predicted direction of the branch long before the result of evaluating the branch condition is available. If the prediction is wrong, the execution of the predicted branch is rolled back, and the other branch is executed. Note that only the architectural state is rolled back, but not the microarchitectural state, which remains visible in contemporary CPUs. Spectre exploits the mismatch between architectural and microarchitectural state by mistraining branch predictors, so victim code (called gadget) executes a mispredicted branch and then rolls back the architectural state, but not the microarchitectural state. The gadget consists of a sequence of instructions, guarded by a conditional branch, that encodes a secret from the victim's own address space into the microarchitectural state, e.g., into the cache by making cache lines cached or not. This encoded information remains accessible to attackers after the misprediction is rolled back by the CPU.

Spectre attacks have so far been demonstrated in JavaScript [Kocher et al., 2019], native code [Kocher et al., 2019, Canella et al., 2019], Go [Duete, 2018] and Rust [Chládek, 2020]. Consequently, a vast variety of mitigations have been proposed, and some of them have been deployed in practice. While Meltdown attacks are mainly mitigated in hardware or microcode, Spectre attacks mainly rely on mitigations on the operating-system level, or in compilers and JIT engines. Given the large number of mitigations and programming languages with associated execution environments, it has become unclear which programming languages, interpreters, compilers and JIT engines, have working mitigations today and can securely be used to implement security-critical code, and which do not.

Spectre does not necessarily work in every single execution environment. The possibility of being able to pull off the attack depends on several factors including (but not limited to) the runtime environment, whether it is fast enough (e.g., interpreters might be too slow for the attack to be pulled off), the compiler and the transformations that it enforces on the code such as optimizations (e.g., whether it emits the code that is efficient enough for this purpose), or the language and its constructs (e.g., do branch-predicted code sequences exit on the binary level). Hence, it was not known which languages, compilers, and environments need mitigations. We are the first to systematically study which languages, compilers, and execution environments need mitigations against Spectre attacks. The result of this study helps developers of languages, compilers, and environments to find out whether or not they should implement mitigations.

In this paper, we systematically analyze the security against Spectre attacks, of the top 30 most-popular programming languages, which involves an analysis of 42 compilers and execution environments. We categorize the languages and their execution environments into three different categories: interpreted, compiled, and managed languages. We determine for each of these languages, compilers and runtimes, whether Spectre mitigations exist and are implemented. The basis of this analysis is publicly available documentation and direct communication with developers. We discover that only 16 of the analyzed execution environments have mitigations for at least one Spectre variant. Based on this initial analysis, we perform a full-scale analysis of the existence of Spectre gadgets for each of these languages and environments. To facilitate our analysis, we present a novel method and tool, *Speconnector*. Speconnector allows us to evaluate and exploit Spectre gadgets independent of the programming language to validate leakage without bi-directional interaction. We use Speconnector to develop Spectre proof-of-concept attacks for 16 programming languages, half of which were previously not known to be affected. Our results highlight that programming languages and their execution environments used to implement security-critical code remain entirely unprotected, even three years after the discovery of the issue, and still continue to leak secrets.

To demonstrate the concrete security impact of our results, we present two case studies that exploit security libraries using Spectre attacks. First, we attack the Alice crypto library, implemented in Java using the OpenJDK compiler. We demonstrate that we can leak the secret key used by Alice for encryption purposes by exploiting a Spectre gadget in it. Second, we attack cryptokit, implemented in OCaml using the ocamlopt native compiler. In this case study, we show that by exploiting a Spectre gadget, we can leak secret data while it resides in memory.

Our work shows that there are still several critical security libraries implemented in programming languages suffering from the lack of mitigations. Security-critical code implemented in these languages remains entirely unprotected against Spectre-type attacks, even three years after the discovery of the issue. We responsibly disclosed our findings to the corresponding vendors and worked with them on solutions. However, we emphasize that this problem goes beyond these specific libraries and is a problem that must be solved by the corresponding compiler, runtime, or interpreter. Finally, until mitigations are available in compilers and runtime environments, developers should be cautious using these programming languages for security-critical code, and users should also be alert using security applications implemented in such languages.

**Contributions.** Our contributions are:

1. A systematic analysis of the security of programming languages and their execution environments with respect to Spectre attacks.
2. We introduce Speconnector, a novel tool to evaluate and exploit Spectre gadgets independent of the programming language.
3. We demonstrate the security impact with two case studies of security-related libraries, and show that we can leak secrets from them.

**Outline.** Section 2 provides background. Section 3 systematically analyzes programming languages for Spectre mitigations. Section 4 presents our new tool Speconnector and our threat model. Section 5 presents our systematic analysis of the Spectre susceptibility of environments using Speconnector. Section 6 presents two case studies showcasing our results. We conclude in Section 7.

# 2 BACKGROUND

In this section, we provide background on speculative and transient execution with a focus on Spectre-type attacks, as well as interpreted, compiled, and managed programming languages.

**Speculative Execution.** Because programs contain conditional branching, CPUs often do not know which instruction needs to be executed next. This occurs, for instance, when the out-of-order execution reaches a not yet completed conditional branch instruction which determines control flow. With speculative execution, the CPU can save the current register state, predict the most likely outcome based on similar events in the past, and *speculatively* execute in the predicted direction. Once the predicted branch instruction has completed execution and the correct path is known, the CPU can decide what to do with the speculatively executed instructions. In the case of a correct prediction, the CPU simply commits (retires) the instructions in the order of the instruction stream, making the results visible on the architectural level. However, if the prediction is not correct, the CPU rolls back to the saved register state, discards the result of the wrongly executed instructions, and continues with the execution of the correct path. The discarded instructions are commonly referred to as *transient instructions* [Kocher, 1996, Lipp et al., 2018, Canella et al., 2019]. While the architectural state is rolled back in the case of a wrong prediction, the microarchitectural state is not reverted, hence allows to infer results and side effects of the transiently executed instructions.

To make such predictions and maximize the performance, CPUs utilize the Branch Prediction Unit (BPU), which consists of different prediction mechanisms [Mittal, 2018]. One such prediction mechanism is the aforementioned conditional branch prediction which is done by the Pattern History Table (PHT) [Canella et al., 2019]. Other mechanisms predict conditional and indirect jump targets [Kocher, 1996] or return addresses of functions [Maisuradze and Rossow, 2018, Koruyeh et al., 2018]. Indirect jumps are predicted by the Branch Target Buffer (BTB) [Lee et al., 2017, Evtyushkin et al., 2016, Kocher et al., 2019], returns by the Return Stack Buffer (RSB) [Maisuradze and Rossow, 2018, Koruyeh et al., 2018].

Branch-prediction logic, e.g., BTB and RSB, is typically not shared across physical cores [Ge et al., 2016], but sometimes among logical cores [Maisuradze and Rossow, 2018]. Therefore, the CPU learns only from previous branches executed on the same physical core.

```
if (x < length_of_data) {
  tmp &= lookup_table[data[x] << 12];
}
```

Figure 1: An example of an index gadget.

**Transient-Execution Attacks.** While speculative and out-of-order execution significantly increase the performance of CPUs [Tomasulo, 1967], these features also have a significant impact on the security of software running on the CPU. So-called transient-execution attacks [Canella et al., 2019] have demonstrated that the effects of transient instructions can be reconstructed on the architectural level as the microarchitectural effects are not reverted. To bring the effects to the architectural level, these types of attacks leverage traditional side-channel attacks, e.g., cache attacks. While the cache has been predominantly exploited for the transmission of the secret data [Kocher et al., 2019, Canella et al., 2019, Maisuradze and Rossow, 2018, Koruyeh et al., 2018], other channels have also been shown to be effective, e.g., execution port contention [Bhattacharyya et al., 2019].

Kocher et al. [Kocher et al., 2019] first discussed transient-execution attacks [Canella et al., 2019] using speculative execution and demonstrated that conditional branches and indirect jumps are potential elements to be exploited to leak data. Subsequent work has then shown that the idea can be extended to function returns [Maisuradze and Rossow, 2018, Koruyeh et al., 2018] and store-to-load forwarding [Horn, 2018]. Canella et al. [Canella et al., 2019] then systematically analyzed the field and demonstrated that the necessary mistraining can be done in the same and a different address space due to some predictors being shared across hyperthreads. Additionally, they showed that many of the proposed countermeasures are ineffective and do not correctly target the root cause of the problem. More recent work [Canella et al., 2020a] has then analyzed the commonalities of these attacks instead of the differences [Canella et al., 2019].

To mitigate all these attacks, various proposals have been made by industry and academia. Canella et al. [Canella et al., 2020b]

analyzed the differences between countermeasures from academia and industry, highlighting that academia proposes more radical countermeasures compared to industry. Proposed mitigations either require hardware changes [Kiriansky et al., 2018, Yan et al., 2018, Khasawneh et al., 2019], secret annotation [Schwarz et al., 2020, Palit et al., 2019, Fustos et al., 2019], new data dependencies [Oleksenko et al., 2018, Carruth, 2018b], or reduced timer accuracy [Pizlo, 2018, Chromium Project, 2018, Wagner, 2018].

**Gadgets.** In transient-execution attacks, a gadget is a piece of code used to transfer the secret information from the victim's side into a covert channel from which the attacker can then retrieve it. Kocher et al. [Kocher et al., 2019] introduced the first Spectre gadget, which is shown in Figure 1. In such a gadget, secret information is stored in memory following the array *data*. A global variable *tmp* is used to prevent optimizing out the memory access. The attacker controls the value of *x*, which allows mistraining the branch for the length check, *i.e.*, the attacker first supplies values for *x* that are lower than *length_of_data*. After mistraining the branch, the attacker supplies a larger value, accesses the secret during transient execution, and encodes it in *lookup_table*, which is shared between victim and attacker. From there, the attacker reads using a cache attack. Canella et al. [Canella et al., 2020a] refer to such a gadget as an *index* gadget.

**Compiled vs Interpreted vs Managed Program Execution.** An important distinction in program execution is between interpretation and compilation. The difference is in whether the program is executed directly using an interpreter, or with an intermediate compilation stage where a compiler generates a new program to be executed. A third option are hybrid compilers where the resulting file of the compilation is again interpreted, combining the other two approaches. We refer to this as *managed program execution.* We often speak of *execution environment* to mean any of compilers, interpreters, or their hybrids.

All of these approaches to execute a program have different advantages and disadvantages. For instance, compiled languages only incur the overhead of translating the code once, while interpreted languages need to translate it every time they are being run. Hence, compilers can perform more sophisticated optimizations as the time needed for translation is less important. LLVM [Lattner and Adve, 2004] is an example of an optimizing compiler framework. C and C++ are examples of programming languages that are usually executed by compilation, although, in principle, they can also be interpreted. An advantage of interpreted languages is that they are more portable as only the interpreter is platform specific, which does not hold for compiled languages. As the interpreter needs to hold more information about the run-time state of the application, interpreted applications typically require more memory during execution than compiled applications. Perl is an example of a languages that is usually interpreted and not compiled.

Hybrid compilers seek to combine the advantages of compiled and interpreted languages. We distinguish several variants of hybrid compilers: Figure 2 shows the two common variants. In the first variant, a compiler first translates the code into platform-independent code, allowing for more sophisticated optimizations, as well as allowing for portability. When running the application, an interpreter executes this platform-independent code. An example of such a hybrid compiler is Ocaml when using ocamlc and ocamlrun: ocamlc carries out the first part to compile the code implemented by the developer, and then ocamlrun interprets the code generated by ocamlc. In the second variant, the compiler again generates platform-independent code. In the second phase, the JIT-compiler uses the platform-independent code to produce the executable code which is interpreted by the interpreter. The interpreter then provides feedback to the JIT-compiler on how to improve the generated code. An example of such hybrid compilers are OracleJDK and C# .NET. Figure 2 provides an illustration of all four aforementioned cases.

# 3 SPECTRE MITIGATIONS BY LANGUAGE AND RUNTIME

When it comes to mitigating Spectre-type attacks, compilers and interpreters play a crucial role as they generate the code that is exploited later on. In this section, we investigate different compilers and interpreters to determine whether they provide methods to mitigate such attacks. Our systematic analysis considers 42 different execution environments which are used to translate the top 30 most-popular programming languages according to GitHub in the third quarter of 2020. The languages encompass compiled, interpreted, and managed languages.

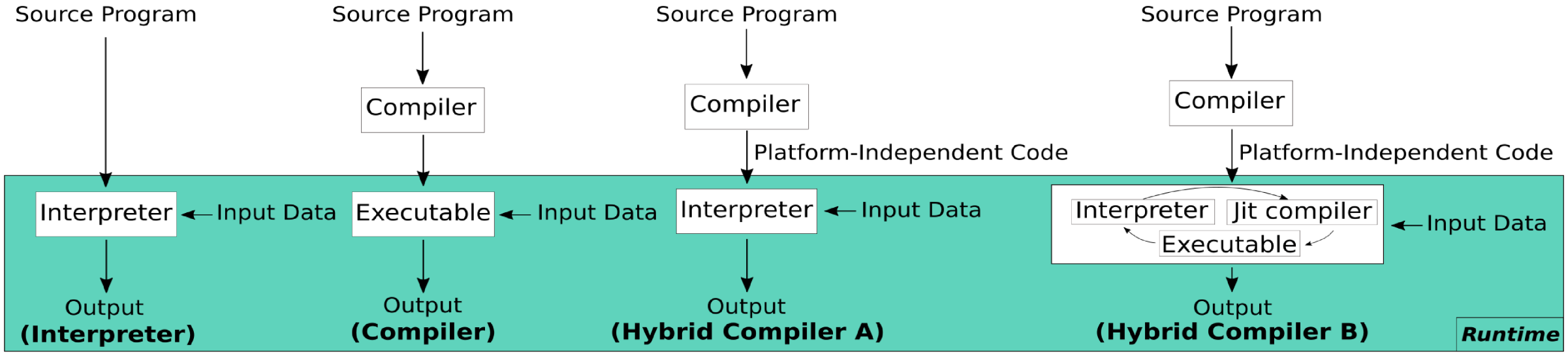

Figure 2: Different categories of programming languages based on their translation process. In this figure, a Perl interpreter is an example of an interpreter, GCC an example of a compiler, OCaml's (ocamlc/ocamlrun) compiler process is an example of a managed language with the translation process in Hybrid Compiler A and OracleJDK shows an example of the translation process for Hybrid Compiler B.

Unfortunately, most of the investigated compilers and interpreters do not provide any published information on mitigations. Hence, as an additional source of information, we contacted developers of the respective compilers and interpreters. Our investigation showed that the majority of the investigated compilers and interpreters did not provide any information whether the lack of Spectre mitigations was because the developers believed their software was safe against Spectre-type attacks. We defer a discussion of the possibility of Spectre-type attacks to Section 5.

Also, based on our investigations, there are several reasons that made the vulnerable execution environments ending up not having any mitigations. One of the main reasons is that language designers mostly do not take Spectre attacks seriously and believe that it is impossible to run the attack using their execution environment, such as in CPython and PyPy. Another reason is that language designers are not aware of the problem in their product, such as in Groovy. For some languages with several layers of compilation, e.g., managed languages, the designers do not believe that mitigations should be provided to software developers on their layer and the problem should be resolved in a another layer, e.g., the Clojure designers believe that this problem should be addressed in JVM. However, this is technically impossible for Spectre-PHT, as the software developer has to explicitly define where to avoid speculation (cf. C's `array_index_nospec`). Ironically, the JDK developers confirm our view, that it is not possible to address the problem in JDK itself [Gross, 2019] without information provided by the software developer in the programming language and environment running on top of JDK. This leaves us in the unfortunate situation that software developers have no possibility to mitigate Spectre when choosing certain programming languages and environments at all. Finally, some of the designers also think that this problem should be addressed at the kernel level, which again is not possible for Spectre-PHT.

Table 1: Interpreted programming languages and their interpreters and which Spectre variants they can mitigate based on the available documentation and direct responses from interpreters' developers. Symbols show whether Spectre mitigations are not available based on the communication with developers (☐), or it was unclear whether there are any mitigations available (X).

| Attack \ PLs | Ruby (MRI) | PHP | Shell (Bash) | Perl | PowerShell (pwsh) | TSQL | Lua | Vim script | Emacs Lisp |
|---|---|---|---|---|---|---|---|---|---|
| Spectre-PHT | X | X | X | ☐ [1] | X | X [2] | X | X | X |
| Spectre-BTB | X | X | X | ☐ [1] | X | X [2] | X | X | X |
| Spectre-RSB | X | X | X | ☐ [1] | X | X [2] | X | X | X |
| Spectre-STL | X | X | X | ☐ [1] | X | X [2] | X | X | X |

[2] [support, 2020]

Below, we briefly describe the process for each category as well as the results of our systematic analysis in more detail, and we rely on the categorization of Spectre-type attacks provided by Canella et al. [Canella et al., 2020a].

**Interpreted Languages.** In this paper, we study 9 different interpreted programming languages and their respective interpreters (Table 1). As a first step in our analysis, we conducted a thorough survey of the publicly available documentation of each interpreter. This step showed that the documentation for 8 of the 9 analyzed interpreters does not contain any information regarding potential mitigations for Spectre. While the documentation for TSQL [support, 2020] contained some information regarding Spectre mitigations, it was unclear whether any mitigations are implemented and which Spectre variants are considered in potential mitigations.

For a more thorough analysis, we analyzed commits made since the public disclosure of Spectre for any indication of available mitigation in the 8 interpreters that do not provide any public information. Unfortunately, this step did not provide any additional insights. Hence, to complete our investigation, we contacted the developers of the respective interpreters to ask for information about potentially implemented mitigations.

In this step, the developers of Perl confirmed that their interpreter does indeed not contain any mitigations for any variant of Spectre [1]. For Ruby (MRI) and PHP, we unfortunately were not able to discuss the issue with the developers. Nevertheless, our results gathered in the first two steps show that we are not aware of any mitigation in these two interpreters. Our analysis for interpreters illustrates that it is unclear for 7 of the 9 analyzed interpreters if they have mitigations implemented to mitigate Spectre attacks

**Compiled Languages.** Our analysis of compiled programming languages shows a different result compared to the analysis of interpreters. In this analysis, we used the same approach as for the interpreted languages and their interpreters, *i.e.*, searched public documentation, analyzed commits, and contacted the developers.

Table 2 presents the results of our analysis. The results show that 13 of the 15 considered compilers provide information about Spectre, but only 12 of them claim to have implemented mitigations for at least one of the existing variants. While all 12 provide mitigations against Spectre-PHT, only 10 provide mitigations for Spectre-BTB. Interestingly, only Go provides a mitigation for Spectre-RSB, while none of the compilers document mitigations against Spectre-STL. In a conversation, the developers of Microsoft's C/C++ compiler confirmed that, despite missing documentation, their compiler provides a Spectre-STL mitigation. The developers of ocamlopt also confirmed in direct communication that their compiler does not provide any form of mitigation against any Spectre variant. Unfortunately, we were not able to get any information about the existence of mitigations for the two remaining compilers, *i.e.*, DM and Haskell (GHC).

Based on our analysis, the Go compiler has the best situation regarding its mitigations against Spectre variants as it contains mitigations for 3 of the 4 Spectre variants discussed by Canella et al. [Canella et al., 2020a].

[1]Dan Book confirmed this via email

[9]Shayne Hiet-Block from Microsoft confirmed this via email

**Managed Languages.** Table 3 shows the results of our analysis of 13 programming languages and their 18 respective hybrid compilers. This analysis provides several interesting insights. For instance, except for the JavaScript compilers SpiderMonkey [Wagner, 2018, de Mooij, 2019], V8 [Chromium Project, 2018, Mcilroy et al., 2019], Chakra [Microsoft, 2018], and the Java compiler GraalVM [Oracle, 2020], none of the other compilers for managed languages provide any form of public documentation of mitigations against at least one of the Spectre variants discussed by Canella et al. [Canella et al., 2020a]. The first three mitigate Spectre-PHT and Spectre-BTB while GraalVM only has mitigations for Spectre-PHT. Similar to the previous case, we again observe that the compilers that do have mitigations, all mitigate Spectre-PHT. While we were able to find documentation regarding Spectre for the Java compiler OpenJDK, the documentation indicates that no mitigations are available.

To counteract the lack of publicly available information on mitigations, we contacted the developers of the PyPy, C#, Scala, Elixir and OCaml (ocamlc/ocamlrun) compilers to gather information on whether they provide mitigations. In all 5 cases, this discussion showed that none of them contain any mitigations. In addition, the documentation for OracleJDK [Oracle, 2021] is hinting at Spectre mitigations, however, it is unclear if any mitigation is implemented and which Spectre variants are the targets in the potential mitigations. Unfortunately, the status of the remaining 7 compilers from this category remain unclear as we did not receive a response. These 7 compilers are Dart, TypeScript, CoffeeScript, Clojure, CPython, Groovy and Kotlin.

## 4 Speconnector

This part describes the goal, tasks, and advantages of the Speconnector tool. Speconnector is written in C, but works for any target language.[30]

The goal of Speconnector is to determine, easily and independent from the target language, whether a given execution environment is vulnerable to Spectre.

[23]Maciej Fijalkowski confirmed this via email

[26]Eric Meadows-Jönsson confirmed this via email

[28]Ilya Pleskunin confirmed this via email

[29]Paul King confirmed this via email

[30]We plan to open source the tool.

Table 2: Compiled programming languages and their compilers and which Spectre variants they can mitigate based on the available documentation and direct responses from the respective compilers' developers. Symbols show whether Spectre mitigations are available based on the documentation (●) or communication with developers (■), not available based on the communication with developers (□), or it was unclear whether there are any mitigations available (X).

| Attack \ PLs | Go | C++ (GCC) | C++ (MS) | C++ (Intel) | C++ (LLVM) | C (GCC) | C (MS) | C (Intel) | C (LLVM) | Rust (LLVM) | Swift (LLVM) | DM | Objective-C (LLVM) | Haskell (GHC) | OCaml (ocamlopt) |
|---|---|---|---|---|---|---|---|---|---|---|---|---|---|---|---|
| Spectre-PHT | ● [3] | ● [5] | ● [8] | ● [10] | ● [12] | ● [5] | ● [8] | ● [10] | ● [12] | ● [12] | ● [12] | X | ● [12] | X | □ [14] |
| Spectre-BTB | ● [3] | ● [6] | □ [9] | ● [10] | ● [13] | ● [6] | □ [9] | ● [10] | ● [13] | ● [13] | ● [13] | X | ● [13] | X | □ [14] |
| Spectre-RSB | ● [3] | ● [7] | □ [9] | □ [11] | X | ● [7] | □ [9] | □ [11] | X | X | X | X | X | X | □ [14] |
| Spectre-STL | □ [4] | X | ■ [9] | □ [11] | X | X | ■ [9] | □ [11] | X | X | X | X | X | X | □ [14] |

[3] [Munday, 2020]
[4] [Randall, 2020]
[5] [GCC.GNU.ORG, 2020]
[6] [Lu, 2018]
[7] [Biener, 2018]
[8] [Pardoe, 2018]
[10] [Jiang, 2018]
[11] [Viet_H_Intel, 2020]
[12] [Carruth, 2020]
[13] [Carruth, 2018a]
[14] [Bacarella, 2020]

Table 3: Managed programming languages and their compilers and which Spectre variants they can mitigate based on the available documentation and direct responses from compilers' developers. Symbols show whether Spectre mitigations are available based on the documentation (●), available but partially effective based on the documentation (◐), not available based on the documentation (○) or communication with developers (□), or it was unclear whether there are any mitigations available (X).

| Attack \ PLs | Dart | Java (OracleJDK) | Java (OpenJDK) | Java (GraalVM) | JavaScript (SpiderMonkey) | JavaScript (V8) | JavaScript (Chakra) | TypeScript | CoffeeScript | Python (PyPy) | C# | Scala | Elixir | Clojure | Python (CPython) | OCaml (ocamlc/ocamlrun) | Kotlin | Groovy |
|---|---|---|---|---|---|---|---|---|---|---|---|---|---|---|---|---|---|---|
| Spectre-PHT | X | X [15] | ○ [16] | ● [17] | ◐ [18,19] | ● [20] | ● [22] | X | X | □ [23] | □ [24] | □ [25] | □ [26] | X | X | □ [27] | □ [28] | □ [29] |
| Spectre-BTB | X | X [15] | ○ [16] | X | ◐ [18,19] | ● [20,21] | ● [22] | X | X | □ [23] | □ [24] | □ [25] | □ [26] | X | X | □ [27] | □ [28] | □ [29] |
| Spectre-RSB | X | X [15] | ○ [16] | X | X | X | X | X | X | □ [23] | □ [24] | □ [25] | □ [26] | X | X | □ [27] | □ [28] | □ [29] |
| Spectre-STL | X | X [15] | ○ [16] | X | X | ○ [21] | X | X | X | □ [23] | □ [24] | □ [25] | □ [26] | X | X | □ [27] | □ [28] | □ [29] |

[15] [Oracle, 2021]
[16] [Gross, 2019]
[17] [Oracle, 2020]
[18] [Wagner, 2018]
[19] [de Mooij, 2019]
[20] [Chromium Project, 2018]
[21] [Mcilroy et al., 2019]
[22] [Microsoft, 2018]
[24] [Tur, 2020]
[25] [Malak and nilskp, 2018]
[27] [Bacarella, 2020]

**Threat Model.**

Speconnector demonstrates that an attack is possible under a regular Spectre attack threat model: The attacker is a co-located program running under the same operating system and is able to execute arbitrary code on the victim machine. The victim code has an interface that we can interact with, e.g., a local API or local server can be a potential target code. Whether an attack is possible under this threat model only depends on whether the victim leaks. Therefore, we focus on the illegal data leakage and use Speconnector to measure and evaluate the victim leakage. While this shows that an attack is possible, crafting a concrete end-to-end exploit for each language requires further engineering steps.

**Algorithm.**

One of the main tasks of Speconnector is to establish shared memory between itself and the victim application. Hence, the target application first allocates 256 pages of memory and fills it with a known magic value while Speconnector also allocates the same amount of memory in its own address space, *i.e.*, the lookup table. Then, Speconnector uses the */proc/[pid]/mem* file of the victim to scan for the pages that contain the magic value. Once it finds them, Speconnector uses the the publicly available PTEditor [Schwarz et al., 2018] to set the victim pages page frame numbers to the page frame numbers of its own lookup table, to establish shared memory between the two processes. As a result, any victim access to one of the now shared pages results in a cache hit when Speconnector performs

Flush+Reload to distinguish between cached and uncached memory.

The advantage of this approach is simplicity: instead re-implementing cache attacks in all programming languages, we can rely on a well understood primitive written in C, e.g., Flush+Reload. For the analysis, the only part that needs to be done in the victim is to implement a victim gadget and to allocate 256 pages and fill them with the known magic value.

# 5 ATTACK FEASIBILITY BY LANGUAGE AND RUNTIME

In this section, we present the results of using Speconnector to determine whether we can establish a covert channel and exploit speculative execution in our set of selected programming languages and their execution environments. To investigate the potential vulnerabilities, we first evaluate each compiler/interpreter to determine whether we can establish a covert channel. Such a covert channel is a necessary building block for Spectre attacks as it gives the attacker the capability of bringing secret data from the microarchitectural state to the architectural level. To determine whether we can establish a covert channel, Speconnector establishes shared memory between itself and the victim application. The victim then accesses the shared lookup table in a known pattern which Speconnector tries to reconstruct. If Speconnector cannot reconstruct the pattern, then it is not possible to establish a covert channel. One potential reason for that can be that the target language uses a different memory layout than C. Then, we evaluate each compiler/interpreter to determine whether we can exploit speculative execution by adding a gadget, mistraining the victim, and using Speconnector to leak the victim's secret data. Because of the nature of the vulnerability and the attack, while a positive result for an execution environment in the experiments shows that the corresponding execution environment is vulnerable and exploitable, a negative result does not guarantee the absence of the vulnerability nor the exploitability.

The experiments were run using different machines for different programming languages and their execution environments. We listed the CPU, the operating system, and the execution environment version for each language and its compiler/interpreter in Table 5 in Appendix A. Table 4 shows the results of our analysis. For completeness, we performed the experiments on the execution environments listed, even if they were very similar, e.g., the both were using JVM in their compilation. The reasoning behind this, is that the construct of each language in its execution environment should be protected against different Spectre variants as well as the composition of different constructs, as certain mitigations, e.g., for Spectre-PHT, can only be implemented on the language level. Therefore, we need to take a look at the problem at the language level.

We now discuss the results in the previously outlined categories of interpreted, compiled, and managed languages individually.

**Interpreted Languages.** Using the approach discussed, we evaluate all our interpreted languages. During this analysis, we were able to exploit Perl's interpreter, successfully demonstrating a covert channel and leaking data using a Spectre attack. The Perl interpreter does not provide any mitigations (cf. Section 3), making it infeasible to protect an application against such attacks. Out of the 9 interpreters we analyzed (cf. Table 4), only the Perl interpreter was vulnerable to both establishing a covert channel and a Spectre attack. A potential explanation is that the speculation window might have been too small for the other interpreters to fit the attack in it. In the remaining 8 interpreters, we were still able to establish a cache covert channel in 5 of them even though we were not able to exploit them with a Spectre attack. We were not able to establish a covert channel in TSQL, Shell (bash) and Vim script. However, this does not mean that these interpreters are safe against Spectre attacks nor against establishing a covert channel. In total, we were able to establish a covert channel in 67 % and exploit 11 % of the analyzed interpreters.

**Compiled Languages.** In our analysis of covert channels in compiled languages, we were able to establish a covert channel in 14 out of 15 (93 %) of the analyzed combinations of languages and compilers. 12 out of the 15 (80 %) compilers generate code that is vulnerable against a Spectre attack. This includes languages and compilers that have been known to be vulnerable but also ones that have so far not been analyzed. One example for such a never before analyzed language and compiler is OCaml with the ocamlopt compiler where we were able to demonstrate both a covert channel and a Spectre attack. While we were able to establish a covert channel in Swift (LLVM) and Haskell (GHC), we were not able to exploit them using a Spectre attack. Compilers annotated with a * in Table 4 are not exploitable or at least less vulnerable in case the available mitigations are used and best practices are followed.

Table 4: Programming languages and their compilers/interpreters and the result whether they are vulnerable to at least one variant of Spectre attack in practice. The first row (depends on settings) shows if compilers/interpreters are not vulnerable (⋆) or are vulnerable (-) if the existing best practices for them are followed.

| Attack \ PLs | Go | C++ (GCC) | C++ (MS) | C++ (Intel) | C++ (LLVM) | C (GCC) | C (MS) | C (Intel) | C (LLVM) | Rust (LLVM) | Swift (LLVM) | DM | Objective-C (LLVM) | Haskell (GHC) | OCaml (ocamlopt) |
|---|---|---|---|---|---|---|---|---|---|---|---|---|---|---|---|
| Depends on setting | ⋆ | ⋆ | ⋆ | ⋆ | ⋆ | ⋆ | ⋆ | ⋆ | ⋆ | ⋆ | ⋆ | - | ⋆ | - | - |
| Covert Channel | ✓ | ✓ | ✓ | ✓ | ✓ | ✓ | ✓ | ✓ | ✓ | ✓ | ✓ | ✗ | ✓ | ✓ | ✓ |
| Spectre Attack | ✓ | ✓ | ✓ | ✓ | ✓ | ✓ | ✓ | ✓ | ✓ | ✓ | ✗ | ✗ | ✓ | ✗ | ✓ |

| Attack \ PLs | Dart | Java (OracleJDK) | Java (OpenJDK) | Java (GraalVM) | JavaScript (SpiderMonkey) | JavaScript (V8) | JavaScript (Chakra) | TypeScript | CoffeeScript | Python (PyPy) | C# | Scala | Elixir | Clojure | Python (CPython) | OCaml (ocamlc/ocamlrun) | Kotlin | Groovy |
|---|---|---|---|---|---|---|---|---|---|---|---|---|---|---|---|---|---|---|
| Depends on setting | - | - | - | ⋆ | ⋆ | ⋆ | ⋆ | - | - | - | - | - | - | - | - | - | - | - |
| Covert Channel | ✓ | ✓ | ✓ | ✓ | ✓ | ✓ | ✓ | ✓ | ✓ | ✓ | ✓ | ✓ | ✓ | ✓ | ✓ | ✓ | ✓ | ✓ |
| Spectre Attack | ✓ | ✓ | ✓ | ✓ | ✓ | ✓ | ✓ | ✓ | ✓ | ✗ | ✓ | ✓ | ✗ | ✗ | ✗ | ✓ | ✓ | ✓ |

| Attack \ PLs | Emacs Lisp | Ruby (MRI) | PHP | Shell (Bash) | Perl | PowerShell (pwsh) | TSQL | Lua | Vim script |
|---|---|---|---|---|---|---|---|---|---|
| Depends on setting | - | - | - | - | - | - | - | - | - |
| Covert Channel | ✓ | ✓ | ✓ | ✗ | ✓ | ✓ | ✗ | ✓ | ✗ |
| Spectre Attack | ✗ | ✗ | ✗ | ✗ | ✓ | ✗ | ✗ | ✗ | ✗ |

**Managed Languages.** For managed languages, we were able to demonstrate a functioning covert channel in all 18 cases (100 %). Out of the 18 language and compiler combinations, we were able to demonstrate a Spectre attack in 14 (78 %) of them. This includes Dart, Java, C#, Scala Groovy, Kotlin and OCaml (ocamlc/ocamlrun), which were so far not known to be vulnerable, *i.e.*, no Spectre attack on these has been demonstrated before. Similar as to compiled languages, there are several instances, again marked with a * in Table 4, where the usage of the available mitigations and following the best practices removes or at least reduces the impact of the vulnerability.

Our investigation of all three language categories provides several interesting insights. First, contrary to interpreted and compiled languages, we were able to establish a covert channel in all analyzed managed languages and compilers, while in the others there was at least one where we were not able to establish a covert channel. Second, compilers for compiled languages are the most vulnerable to Spectre attacks if mitigations are not used and best practices are not followed. Third, in all three categories, we were able to extract data using a Spectre attack in at least one instance. This further highlights that many languages and their respective compilers/interpreters have not considered Spectre attacks even 3 years after they were introduced to the public.

# 6 CASE STUDIES

In this section, we present two case studies, each using one of the vulnerable programming languages taken from Table 4. For both, we demonstrate how a Spectre attack can be used to leak secret information from real-world libraries. Furthermore, we argue that a mitigation at the compiler level prevents our attacks.

## 6.1 Alice - Java

In our first case study, we investigate *Alice*[8], a crypto library written in Java which provides encryption/decryption with AES of byte arrays and files. Alice supports different key sizes, block modes and padding schemes. Our goal is to find and exploit speculative execution vulnerabilities to leak the secret key which is being used to encrypt/decrypt the plaintext in Alice encryption and decryption methods.

We first performed an in-depth assessment of this library and the way that it should be used. The library contains three main Java classes. Class *Alice* which is the main API in the library for encryption purposes. Class *AliceContext* which provides the context to be passed to class *Alice* to create a new object of it. And class *AliceContextBuilder* which is used to build the context. More information about Alice library is provided on their publicly available documents.

Our attack uses three Java classes, *Main*, *Enc* and *Victim*. The *Main* class is responsible for starting two threads on the same physical core, one uses *Enc* while the other uses *Victim*. *Enc* creates an instance of an unmodified *Alice* and performs the encryption using the context generated by *AliceContextBuilder* and the password stored within *Enc*. Our *Victim* class contains an index gadget (cf. Figure 1) mistrained and used by the attacker to leak the secret information.

As a first step in our attack, our *Main* class starts both the encryption and the victim containing the gadget. Once this is done, we use

[8]https://rockaport.github.io/alice/

Speconnector (cf. Section 4) simply to ease the exploitation, but note that extracting the data is also possible directly in Java. Speconnector establishes a shared memory with the victim by allocating 256 pages and filling them with the known magic value. The victim is mistrained to perform the speculative out-of-bounds access which causes a secret-dependent cache access on our shared memory. To extract the password stored in *Enc*, Speconnector constantly performs Flush+Reload on the memory shared with the victim. As OpenJDK does not contain any Spectre mitigations (cf. Table 3), Speconnector can successfully extract the password, compromising the whole encryption system.

## 6.2 cryptokit - OCaml

Our second case study performs a thorough analysis of *cryptokit* [Leroy, 2020], a widely used OCaml crypto library. As Tables 2 and 3 show, the documentation of OCaml already indicates that no mitigations are available against Spectre attacks. It is an important case study, as so far OCaml has not been shown to be affected.

In this analysis, we encountered several factors that complicate the analysis. First, OCaml can be either compiled into native machine code using the ocamlopt compiler, or it can be compiled into bytecode using the ocamlc compiler, which is then executed by the ocamlrun runtime. Depending on the compiler being used, it can either be regarded as a compiled or a managed language. Second, the environment is important as well since some gadgets show themselves only in some particular configurations. For example, while we could not cause any data leakage with our Spectre index gadget when using the ocamlc bytecode compiler and the ocamlrun runtime on an older Intel Core2Duo T9600 CPU, the exact same gadget worked well when run on a more modern Intel Core I7-6700K CPU. When compiled to native code using the ocamlopt compiler, the same source code produces a gadget that works on both machines. For best reproducibility, we therefore only used the ocamlopt native compiler in the following case study.

The cryptokit library implements a variety of cryptographic primitives for use in security-sensitive applications, including symmetric-key ciphers, public-key cryptography, and hash functions. Our manual analysis has not shown an exploitable Spectre gadget within the library itself, but the possibility exists that the application that uses it leaks the sensitive data processed by the library by executing a gadget while this data is directly used or parts of it remain in memory. For example, cryptokit has a function `transform_string t s`, which applies the given transform object `t` to the given string `s`. The transform object itself describes the desired cryptographic action, including its parameters. For AES encryption, these include the key and the initialization vector (IV). After the transformation, the internal state of the transform object `t` is wiped, but this does not include the key and the IV. Furthermore, the string `s` is not wiped, either. The reason is that, in OCaml, strings are passed by reference across function calls. This makes it unsafe for the library function to overwrite these strings as the calling code might still need them. Therefore, the application itself is responsible for wiping the key, the IV and the plaintext. However, it is worth noting that a direct hint about this is missing in the current documentation (version 1.16.1) of the library [Leroy, 2020] even though it offers a function `wipe_string s` for this purpose. Furthermore, the official OCaml documentation even states that strings should be treated as being immutable and advises against overwriting them [Leroy et al., 2020].

To verify and substantiate this finding, we performed the following experiment. Following a best-practice example of using cryptokit, we wrote a proof-of-concept application that encrypts user-provided plaintext. Similar to the Java case study in the previous section, we allocate 256 contiguous pages using OCaml's *Array1* module and fill them with the known magic value. Then we add a Spectre gadget to the application. Using that gadget to perform speculative out-of-bounds accesses on a string and connecting Speconnector to it, we were able to successfully reconstruct the original plaintext after encrypting it and after the last reference to the plaintext went out of scope. To find the actual location of the plaintext in memory, we computed the required offset for the out-of-bounds access in the proof-of-concept application itself. This showed that the offset is the same each time the program is run, making it predictable for the attacker. Obtaining the key or the IV instead of the plaintext should also be possible using the same approach. By this, we show that an OCaml application can leak sensitive data via transient execution side-channels as long as that data is in memory. As neither the native compiler (ocamlopt) nor the bytecode compiler (ocamlopt) and runtime (ocamlrun) provide any form of mitigations, it is the task of the application developer to ensure that no Spectre gadget is executed while sensitive data resides in memory. To wipe sensitive data from memory, the programmer has to overwrite strings which are otherwise considered to be immutable by the language.

Furthermore, sensitive data can also remain in internal input or output buffers, which cannot be overwritten by the application developer. For example, to read files, OCaml uses channel objects, which are buffered [OCAML.ORG, 2021] yet provide no means of wiping that buffer [Leroy et al., 2020]. We verified that sensitive data can indeed be leaked from such buffers by modifying our experiment to leak the data directly from the input buffer of the standard input channel. Again, we computed the offset for the out-of-bounds access in the victim. We observed that Address Space Layout Randomization (ASLR) leads to different offsets each time the program is run, because the channel buffer is stored in another memory mapping than the OCaml strings. However, the offset of the buffer within its memory mapping itself is constant. Turning off ASLR results in a constant offset between the OCaml string and the channel buffer.

In conclusion, sensitive data might live longer in memory than the developer might expect and, therefore, also be susceptible to Spectre attacks for quite a long time. Hence, it is imperative that OCaml provides ways for a developer to either reliably remove data from sensitive buffers or to hinder transient execution at all.

# 7 CONCLUSION

We presented a systematic analysis of different programming languages and their respective compilers/interpreters against Spectre attacks. Our analysis uncovered Spectre attacks in 16 different programming languages with 8 of them not investigated so far and not known to be vulnerable. We also demonstrated practical proof-of-concept attacks and evaluated their applicability to these programming languages and their respective execution environments. With Speconnector, we presented a novel approach and tool that can be used to quickly verify whether a programming language and execution environment is vulnerable to Spectre attacks. We further presented two case studies of real world libraries in two different programming languages, demonstrating that they are susceptible to Spectre attacks. Our results indicate that it is necessary to do more research into mitigations that target a wide range of programming languages and execution environments.

# REFERENCES

[Bacarella, 2020] Bacarella, M. (2020). Spectre mitigations in ocaml compiler?!

[Bhattacharyya et al., 2019] Bhattacharyya, A., Sandulescu, A., Neugschwandt ner, M., Sorniotti, A., Falsafi, B., Payer, M., and Kurmus, A. (2019). SMoTherSpectre: exploiting speculative execution through port contention. In *CCS*.

[Biener, 2018] Biener, R. (2018). GCC 7.3 released.

[Canella et al., 2020a] Canella, C., Khasawneh, K. N., and Gruss, D. (2020a). The Evolution of Transient-Execution Attacks. In *GLSVLSI*.

[Canella et al., 2020b] Canella, C., Pudukotai Dinakarrao, S. M., Gruss, D., and Khasawneh, K. N. (2020b). Evolution of Defenses against Transient-Execution Attacks. In *GLSVLSI*.

[Canella et al., 2019] Canella, C., Van Bulck, J., Schwarz, M., Lipp, M., von Berg, B., Ortner, P., Piessens, F., Evtyushkin, D., and Gruss, D. (2019). A Systematic Evaluation of Transient Execution Attacks and Defenses. In *USENIX Security Symposium*. Extended classification tree and PoCs at https://transient.fail/.

[Carruth, 2018a] Carruth, C. (2018a).

[Carruth, 2018b] Carruth, C. (2018b). RFC: Speculative Load Hardening (a Spectre variant #1 mitigation).

[Carruth, 2020] Carruth, C. (2020). Speculative load hardening.

[Chládek, 2020] Chládek, J. (2020). rust-spectre.

[Chromium Project, 2018] Chromium Project (2018). Mitigating Side-Channel Attacks.

[de Mooij, 2019] de Mooij, J. (2019). [Meta] Spectre JIT mitigations.

[Duete, 2018] Duete, L. (2018). spectre-go.

[Evtyushkin et al., 2016] Evtyushkin, D., Ponomarev, D., and Abu-Ghazaleh, N. (2016). Jump over ASLR: Attacking branch predictors to bypass ASLR. In *MICRO*.

[Fustos et al., 2019] Fustos, J., Farshchi, F., and Yun, H. (2019). SpectreGuard: An Efficient Data-centric Defense Mechanism against Spectre Attacks. In *DAC*.

[GCC.GNU.ORG, 2020] GCC.GNU.ORG (2020). Other built-in functions provided by gcc.

[Ge et al., 2016] Ge, Q., Yarom, Y., Cock, D., and Heiser, G. (2016). A Survey of Microarchitectural

Timing Attacks and Countermeasures on Contemporary Hardware. *Journal of Cryptographic Engineering*.

[Gross, 2019] Gross, A. (2019). Java and speculative execution vulnerabilities.

[Horn, 2018] Horn, J. (2018). speculative execution, variant 4: speculative store bypass.

[Jiang, 2018] Jiang, J. L. (2018). Using intel® compilers to mitigate speculative execution side-channel issues.

[Khasawneh et al., 2019] Khasawneh, K. N., Koruyeh, E. M., Song, C., Evtyushkin, D., Ponomarev, D., and Abu-Ghazaleh, N. (2019). SafeSpec: Banishing the Spectre of a Meltdown with Leakage-Free Speculation. In *DAC*.

[Kiriansky et al., 2018] Kiriansky, V., Lebedev, I., Amarasinghe, S., Devadas, S., and Emer, J. (2018). DAWG: A Defense Against Cache Timing Attacks in Speculative Execution Processors. In *MICRO*.

[Kocher et al., 2019] Kocher, P., Horn, J., Fogh, A., Genkin, D., Gruss, D., Haas, W., Hamburg, M., Lipp, M., Mangard, S., Prescher, T., Schwarz, M., and Yarom, Y. (2019). Spectre Attacks: Exploiting Speculative Execution. In *S&P*.

[Kocher, 1996] Kocher, P. C. (1996). Timing Attacks on Implementations of Diffe-Hellman, RSA, DSS, and Other Systems. In *CRYPTO*.

[Koruyeh et al., 2018] Koruyeh, E. M., Khasawneh, K., Song, C., and Abu-Ghazaleh, N. (2018). Spectre Returns! Speculation Attacks using the Return Stack Buffer. In *WOOT*.

[Lattner and Adve, 2004] Lattner, C. and Adve, V. S. (2004). LLVM: A compilation framework for lifelong program analysis & transformation. In *IEEE / ACM International Symposium on Code Generation and Optimization – CGO*.

[Lee et al., 2017] Lee, S., Shih, M., Gera, P., Kim, T., Kim, H., and Peinado, M. (2017). Inferring Fine-grained Control Flow Inside SGX Enclaves with Branch Shadowing. In *USENIX Security Symposium*.

[Leroy, 2020] Leroy, X. (2020). cryptokit/cryptokit.mli at release1161.

[Leroy et al., 2020] Leroy, X., Doligez, D., Frisch, A., Garrigue, J., Rémy, D., and Vouillon, J. (2020). The OCaml system release 4.11 – Documentation and user's manual.

[Lipp et al., 2018] Lipp, M., Schwarz, M., Gruss, D., Prescher, T., Haas, W., Fogh, A., Horn, J., Mangard, S., Kocher, P., Genkin, D., Yarom, Y., and Hamburg, M. (2018). Meltdown: Reading Kernel Memory from User Space. In *USENIX Security Symposium*.

[Lu, 2018] Lu, H. (2018). x86: Cve-2017-5715, aka spectre.

[Maisuradze and Rossow, 2018] Maisuradze, G. and Rossow, C. (2018). ret2spec: Speculative Execution Using Return Stack Buffers. In *CCS*.

[Malak and nilskp, 2018] Malak, M. and nilskp (2018). Meltdown/spectre.

[Mcilroy et al., 2019] Mcilroy, R., Sevcik, J., Tebbi, T., Titzer, B. L., and Verwaest, T. (2019). Spectre is here to stay: An analysis of side-channels and speculative execution. *arXiv:1902.05178*.

[Microsoft, 2018] Microsoft (2018). Mitigating speculative execution side-channel attacks in Microsoft Edge and Internet Explorer.

[Mittal, 2018] Mittal, S. (2018). A Survey of Techniques for Dynamic Branch Prediction. *arxiv:1804.00261*.

[Munday, 2020] Munday, M. (2020). consider extending '-spectre' option to other architectures.

[OCAML.ORG, 2021] OCAML.ORG (2021). File manipulation.

[Oleksenko et al., 2018] Oleksenko, O., Trach, B., Reiher, T., Silberstein, M., and Fetzer, C. (2018). You Shall Not Bypass: Employing data dependencies to prevent Bounds Check Bypass. *arXiv:1805.08506*.

[Oracle, 2020] Oracle (2020). Oracle graalvm enterprise edition 20 guide.

[Oracle, 2021] Oracle (2021). Oracle JVM Security Enhancements.

[Palit et al., 2019] Palit, T., Monrose, F., and Polychronakis, M. (2019). Mitigating data leakage by protecting memory-resident sensitive data. In *ACSAC*.

[Pardoe, 2018] Pardoe, A. (2018). Spectre mitigations in msvc.

[Pizlo, 2018] Pizlo, F. (2018). What Spectre and Meltdown mean for WebKit.

[Randall, 2020] Randall, K. (2020). consider extending '-spectre' option to other architectures.

[Schwarz et al., 2018] Schwarz, M., Lipp, M., and Canella, C. (2018). misc0110/PTEditor: A small library to modify all page-table levels of all processes from user space for x86_64 and ARMv8.

[Schwarz et al., 2020] Schwarz, M., Lipp, M., Canella, C., Schilling, R., Kargl, F., and Gruss, D. (2020). ConTExT: A Generic Approach for Mitigating Spectre. In *NDSS*.

[support, 2020] support, M. (2020). Kb4073225 - sql server guidance to protect against spectre, meltdown and micro-architectural data sampling vulnerabilities.

[Tomasulo, 1967] Tomasulo, R. M. (1967). An efficient algorithm for exploiting multiple arithmetic units. *IBM Journal of research and Development*, 11(1):25–33.

[Tur, 2020] Tur, J. (2020). Consider porting interlocked.speculationbarrier() from .net 4.8.

[Viet_H_Intel, 2020] Viet_H_Intel (2020). Spectre mitigations in intel c++ compiler.

[Wagner, 2018] Wagner, L. (2018). Mitigations landing for new class of timing attack.

[Yan et al., 2018] Yan, M., Choi, J., Skarlatos, D., Morrison, A., Fletcher, C. W., and Torrellas, J. (2018). InvisiSpec: Making Speculative Execution Invisible in the Cache Hierarchy. In *MICRO*.

# A EXPERIMENT SETUP

Table 5 summarizes the setups of our experiments.

Table 5: Programming languages and their compilers/interpreters and the CPU, operating system, and the execution environment which we used for our experiments. The first row (source of attack) shows if we have performed the attack (●), we have not tried but other have done it (○), we were not able to perform the attack (X). The next row (depends on settings) shows if compilers/interpreters are not vulnerable (⋆) or are vulnerable (-) if the existing best practices for them are followed. A checkmark (✓) shows that it was used for an experiment while a dash (-) indicates that the respective language was not tested on this CPU or OS. For the execution environment, the dash (-) indicates that it does not apply to the respective language and compiler/interpreter.

| | Attack \ PLs | Go | C++ (GCC) | C++ (MS) | C++ (Intel) | C++ (LLVM) | C (GCC) | C (MS) | C (Intel) | C (LLVM) | Rust (LLVM) | Kotlin | Swift (LLVM) | Groovy | DM | Objective-C (LLVM) | Haskell (GHC) | OCaml (ocamlopt) | Dart | Java (OracleJDK) | Java (OpenJDK) | Java (GraalVM) | JavaScript (SpiderMonkey) | JavaScript (V8) | JavaScript (Chakra) | TypeScript | CoffeeScript | Python (PyPy) | C# | Scala | Elixir | Clojure | Python (CPython) | OCaml (ocamlc/ocamlrun) | Emacs Lisp | Ruby (MRI) | PHP | Shell (Bash) | Perl | PowerShell (pwsh) | TSQL | Lua | Vim script |
|---|---|---|---|---|---|---|---|---|---|---|---|---|---|---|---|---|---|---|---|---|---|---|---|---|---|---|---|---|---|---|---|---|---|---|---|---|---|---|---|---|---|---|---|
| | Source of attack | ○ | ● | ○ | ● | ● | ● | ○ | ● | ● | ○ | ● | ● | ● | X | ● | ● | ● | ● | ● | ● | ● | ○ | ○ | ○ | ○ | ○ | ● | ● | ● | ● | ● | ● | ● | ● | ● | ● | ● | ● | ● | ● | ● | ● |
| | Depends on setting | ⋆ | ⋆ | ⋆ | ⋆ | ⋆ | ⋆ | ⋆ | ⋆ | ⋆ | ⋆ | - | ⋆ | - | - | ⋆ | - | - | - | - | - | ⋆ | ⋆ | ⋆ | ⋆ | - | - | - | - | - | - | - | - | - | - | - | - | - | - | - | - | - | - |
| CPU | i7-2630QM | - | ✓ | - | - | - | ✓ | - | - | - | - | ✓ | - | ✓ | - | - | ✓ | - | - | ✓ | ✓ | ✓ | - | - | - | - | - | - | - | ✓ | - | ✓ | - | - | - | - | - | - | ✓ | - | - | - | - |
| | i5-6200U | - | ✓ | - | ✓ | ✓ | ✓ | - | ✓ | ✓ | - | ✓ | - | ✓ | - | ✓ | ✓ | - | - | ✓ | ✓ | ✓ | - | - | - | - | - | - | - | ✓ | - | ✓ | - | - | - | - | - | - | - | - | ✓ | - | - |
| | i7-6700K | - | - | - | - | - | - | - | - | - | - | - | ✓ | - | - | - | - | ✓ | ✓ | - | - | - | - | - | - | - | - | ✓ | - | - | ✓ | - | ✓ | ✓ | ✓ | ✓ | ✓ | ✓ | - | - | - | ✓ | ✓ |
| | Xeon Silver 4208 | - | - | - | - | - | - | - | - | - | - | - | - | - | - | - | - | - | - | - | - | - | - | - | - | - | - | - | ✓ | - | - | - | - | - | - | - | - | - | - | ✓ | - | - | - |
| OS | Ubuntu 20.4 LTS | - | ✓ | - | ✓ | ✓ | ✓ | - | ✓ | ✓ | - | ✓ | ✓ | ✓ | - | ✓ | ✓ | ✓ | ✓ | ✓ | ✓ | ✓ | - | - | - | - | - | ✓ | ✓ | ✓ | ✓ | ✓ | ✓ | ✓ | ✓ | ✓ | ✓ | ✓ | ✓ | ✓ | ✓ | ✓ | ✓ |
| Environment | gcc 9.3.0 | - | ✓ | - | - | - | ✓ | - | - | - | - | - | - | - | - | ✓ | - | - | - | - | - | - | - | - | - | - | - | - | - | - | - | - | - | - | - | - | - | - | - | - | - | - | - |
| | IntelCompiler 2021.1.2 | - | - | - | ✓ | - | - | - | ✓ | - | - | - | - | - | - | - | - | - | - | - | - | - | - | - | - | - | - | - | - | - | - | - | - | - | - | - | - | - | - | - | - | - | - |
| | clang 12.0.0 | - | - | - | - | ✓ | - | - | - | ✓ | - | - | - | - | - | - | - | - | - | - | - | - | - | - | - | - | - | - | - | - | - | - | - | - | - | - | - | - | - | - | - | - | - |
| | kotlinc-jvm 1.4.20 (JRE 11.0.9.1) | - | - | - | - | - | - | - | - | - | - | ✓ | - | - | - | - | - | - | - | - | - | - | - | - | - | - | - | - | - | - | - | - | - | - | - | - | - | - | - | - | - | - | - |
| | Swift 5.3.1 | - | - | - | - | - | - | - | - | - | - | - | ✓ | - | - | - | - | - | - | - | - | - | - | - | - | - | - | - | - | - | - | - | - | - | - | - | - | - | - | - | - | - | - |
| | Groovy 3.0.4 | - | - | - | - | - | - | - | - | - | - | - | - | ✓ | - | - | - | - | - | - | - | - | - | - | - | - | - | - | - | - | - | - | - | - | - | - | - | - | - | - | - | - | - |
| | GHC 8.6.5 | - | - | - | - | - | - | - | - | - | - | - | - | - | - | - | ✓ | - | - | - | - | - | - | - | - | - | - | - | - | - | - | - | - | - | - | - | - | - | - | - | - | - | - |
| | ocamlopt 4.08.1 | - | - | - | - | - | - | - | - | - | - | - | - | - | - | - | - | ✓ | - | - | - | - | - | - | - | - | - | - | - | - | - | - | - | - | - | - | - | - | - | - | - | - | - |
| | Dart SDK version: 2.10.5 (stable) | - | - | - | - | - | - | - | - | - | - | - | - | - | - | - | - | - | ✓ | - | - | - | - | - | - | - | - | - | - | - | - | - | - | - | - | - | - | - | - | - | - | - | - |
| | JAVA 11.0.9 | - | - | - | - | - | - | - | - | - | - | - | - | - | - | - | - | - | - | ✓ | ✓ | - | - | - | - | - | - | - | - | - | - | - | - | - | - | - | - | - | - | - | - | - | - |
| | GRAALVM 20.3.0 | - | - | - | - | - | - | - | - | - | - | - | - | - | - | - | - | - | - | - | - | ✓ | - | - | - | - | - | - | - | - | - | - | - | - | - | - | - | - | - | - | - | - | - |
| | PyPy 7.3.1 | - | - | - | - | - | - | - | - | - | - | - | - | - | - | - | - | - | - | - | - | - | - | - | - | - | - | ✓ | - | - | - | - | - | - | - | - | - | - | - | - | - | - | - |
| | .Net Core (dotnet) 5.0.100 | - | - | - | - | - | - | - | - | - | - | - | - | - | - | - | - | - | - | - | - | - | - | - | - | - | - | - | ✓ | - | - | - | - | - | - | - | - | - | - | - | - | - | - |
| | Scala 2.13.4 | - | - | - | - | - | - | - | - | - | - | - | - | - | - | - | - | - | - | - | - | - | - | - | - | - | - | - | - | ✓ | - | - | - | - | - | - | - | - | - | - | - | - | - |
| | Elixir - IEx 1.9.1 | - | - | - | - | - | - | - | - | - | - | - | - | - | - | - | - | - | - | - | - | - | - | - | - | - | - | - | - | - | ✓ | - | - | - | - | - | - | - | - | - | - | - | - |
| | Clojure 1.10.1 | - | - | - | - | - | - | - | - | - | - | - | - | - | - | - | - | - | - | - | - | - | - | - | - | - | - | - | - | - | - | ✓ | - | - | - | - | - | - | - | - | - | - | - |
| | CPython 3.8.2 | - | - | - | - | - | - | - | - | - | - | - | - | - | - | - | - | - | - | - | - | - | - | - | - | - | - | - | - | - | - | - | ✓ | - | - | - | - | - | - | - | - | - | - |
| | ocamlc/ocamlrun 4.08.1 | - | - | - | - | - | - | - | - | - | - | - | - | - | - | - | - | - | - | - | - | - | - | - | - | - | - | - | - | - | - | - | - | ✓ | - | - | - | - | - | - | - | - | - |
| | GNU Emacs 26.3 | - | - | - | - | - | - | - | - | - | - | - | - | - | - | - | - | - | - | - | - | - | - | - | - | - | - | - | - | - | - | - | - | - | ✓ | - | - | - | - | - | - | - | - |
| | ruby 2.7.0 | - | - | - | - | - | - | - | - | - | - | - | - | - | - | - | - | - | - | - | - | - | - | - | - | - | - | - | - | - | - | - | - | - | - | ✓ | - | - | - | - | - | - | - |
| | PHP 7.4.3 | - | - | - | - | - | - | - | - | - | - | - | - | - | - | - | - | - | - | - | - | - | - | - | - | - | - | - | - | - | - | - | - | - | - | - | ✓ | - | - | - | - | - | - |
| | bash 5.0.17(1)-release | - | - | - | - | - | - | - | - | - | - | - | - | - | - | - | - | - | - | - | - | - | - | - | - | - | - | - | - | - | - | - | - | - | - | - | - | ✓ | - | - | - | - | - |
| | perl v5.30.0 | - | - | - | - | - | - | - | - | - | - | - | - | - | - | - | - | - | - | - | - | - | - | - | - | - | - | - | - | - | - | - | - | - | - | - | - | - | ✓ | - | - | - | - |
| | PowerShell 7.1.0 | - | - | - | - | - | - | - | - | - | - | - | - | - | - | - | - | - | - | - | - | - | - | - | - | - | - | - | - | - | - | - | - | - | - | - | - | - | - | ✓ | - | - | - |
| | SQL Server Command Line Tool Version 17.6 | - | - | - | - | - | - | - | - | - | - | - | - | - | - | - | - | - | - | - | - | - | - | - | - | - | - | - | - | - | - | - | - | - | - | - | - | - | - | - | ✓ | - | - |
| | Lua 5.4.1 | - | - | - | - | - | - | - | - | - | - | - | - | - | - | - | - | - | - | - | - | - | - | - | - | - | - | - | - | - | - | - | - | - | - | - | - | - | - | - | - | ✓ | - |
| | vim 8.1-2269 | - | - | - | - | - | - | - | - | - | - | - | - | - | - | - | - | - | - | - | - | - | - | - | - | - | - | - | - | - | - | - | - | - | - | - | - | - | - | - | - | - | ✓ |